\documentclass[runningheads]{llncs}
\usepackage[T1]{fontenc}
\usepackage{graphicx}
\usepackage{nicematrix} 
\usepackage{enumitem} 
\setlist{parsep=0pt,listparindent=\parindent}
\usepackage{hyperref} 
\usepackage{booktabs}  
\usepackage{siunitx}   
\usepackage{amsmath} 
\usepackage[misc]{ifsym} 
\begin{document}

\title{Towards Enhancing Linked Data Retrieval in Conversational UIs using Large Language Models}
\titlerunning{Towards Enhancing LD Retrieval in Conversational UIs using LLMs}

\author{Omar Mussa\inst{1,3}\textsuperscript{\Letter}\orcidID{0000-0001-8614-6550} \and
Omer Rana\inst{1}\orcidID{0000-0003-3597-2646} \and
Beno\^{\i}t Goossens\inst{2}\orcidID{0000-0003-2360-4643} \and
Pablo Orozco-terWengel\inst{2}\orcidID{0000-0002-7951-4148} \and
Charith Perera\inst{1}\orcidID{0000-0002-0190-3346}}

\authorrunning{O. Mussa et al.}

\institute{School of Computer Science and Informatics, Cardiff University, Cardiff, UK
\email{\{MussaO,RanaOF,PereraC\}@cardiff.ac.uk}
\and
School of Biosciences, Cardiff University, Cardiff, UK\\
\email{\{GoossensBR,Orozco-terWengelPA\}@cardiff.ac.uk}
\and
College of Computing and Informatics, Saudi Electronic University, Riyadh, KSA\\
\email{o.mousa@seu.edu.sa}
}

\maketitle              
\begin{abstract}
Despite the recent broad adoption of Large Language Models (LLMs) across various domains, their potential for enriching information systems in extracting and exploring Linked Data (LD) and Resource Description Framework (RDF) triplestores has not been extensively explored. This paper examines the integration of LLMs within existing systems, emphasising the enhancement of conversational user interfaces (UIs) and their capabilities for data extraction by producing more accurate SPARQL queries without the requirement for model retraining. Typically, conversational UI models necessitate retraining with the introduction of new datasets or updates, limiting their functionality as general-purpose extraction tools. Our approach addresses this limitation by incorporating LLMs into the conversational UI workflow, significantly enhancing their ability to comprehend and process user queries effectively. By leveraging the advanced natural language understanding capabilities of LLMs, our method improves RDF entity extraction within web systems employing conventional chatbots. This integration facilitates a more nuanced and context-aware interaction model, critical for handling the complex query patterns often encountered in RDF datasets and Linked Open Data (LOD) endpoints. The evaluation of this methodology shows a marked enhancement in system expressivity and the accuracy of responses to user queries, indicating a promising direction for future research in this area. This investigation not only underscores the versatility of LLMs in enhancing existing information systems but also sets the stage for further explorations into their potential applications within more specialised domains of web information systems.

\keywords{Conversational UIs  \and Large Language Models \and Linked Data \and RDF Triplestore \and Information Retrieval.}
\end{abstract}

\section{Introduction}
In recent years, significant advancements have been made in the development of large language models (LLMs), which have shown robust capabilities across a broad spectrum of natural language processing (NLP) tasks. These models can execute tasks without the need for specific fine-tuning, as they can be directed through instructional prompts embedded within the requests, thereby facilitating their application across various domains~\cite{Zhang2022b,Chung2024}. This scientific progress has catalysed a surge in research aimed at examining their applicability and efficacy across a diverse array of topics and methodologies such as education~\cite{Baidoo2023}, healthcare~\cite{Sallam2023}, and finance~\cite{Dowling2023}, as well as for specific NLP tasks like replacing traditional question-answering models~\cite{Tan2023,Feng2023}. In particular, commercial versions of these models, such as GPT, have received considerable attention and have been promoted as optimal solutions for integration within chatbot applications and backend text processing solutions. This has opened up new avenues for leveraging artificial intelligence to enhance user interactions and streamline data processing.

Despite these advancements, one area that still needs to be explored is the potential of LLMs to serve as entity extractors within existing linked data (LD) systems and Resource Description Framework (RDF) triplestores. The integration of LLMs in such capacities could potentially revolutionise the way entities are extracted and managed, offering more dynamic and context-aware data handling capabilities. However, the efficacy and reliability of LLMs in this specific role have not been thoroughly tested, raising questions about their practicality and performance in real-world settings.

In this study, we employ a novel toolkit known as ForestQB~\cite{Mussa2024}, specifically developed to explore observational LD to support bioscientists and wildlife conservation efforts. This system integrates an interface that combines a chatbot with a traditional form-based graphical user interface (GUI). It facilitates the extraction of data from diverse observational LD endpoints. The chatbot within this toolkit is configured to interpret user queries and handle entity extraction. This process automates populating the GUI with appropriate selections and executes the search process. Consequently, the chatbot, as a general-purpose tool, is designed to adapt to changes in the dataset, thereby not being restricted to specific names or sentences within its training data in an ontology-independent manner. Therefore, the model used by the chatbot lacks the ability to comprehend the relationships within the knowledge graph concerning the sensors and their observations. As a result, constructing queries that incorporate an arbitrary array of filters and entities through the Conversational UI is unfeasible. In addition, general queries about the descriptions of the sensors and entities within the dataset cannot be accommodated.

The objective of our research is to critically evaluate the validity and effectiveness of implementing this approach within current systems. Our study aims to explore how the integration of LLMs can enhance the expressivity and functionality of systems dealing with LD and RDF triplestores. By conducting empirical tests and analysing the outcomes, we hope to provide insights into the benefits and limitations of using LLMs as entity extractors, thus contributing to the broader discourse on the applicability of LLMs in complex data environments. Our contribution are summarised as follows:
\begin{itemize}
    \item We introduce a novel strategy for enhancing entity extraction and question answering over observational linked data by utilising LLMs as a pipeline component in conversational UIs systems.

    \item A comprehensive set of design principles to assist practitioners in enhancing their research through the integration of LLMs.
    
    \item We conducted empirical testing with state-of-the-art LLMs to assess the efficacy of our methodology. The findings indicate a significant potential to improve information retrieval and reasoning processes over RDF data.
\end{itemize}

\section{Related Work}
\paragraph{\textbf{Entity Extraction.}} Traditional entity extraction, which identifies and classifies named entities within unstructured text, relies on rule-based systems and machine learning algorithms. These methods depend on predefined patterns, dictionaries and statistical models but are hindered by the need for frequent updates and extensive annotated datasets \cite{Hu2024}. They also struggle with context-specific nuances, limiting their ability to integrate additional reference data to improve accuracy \cite{Hu2024}. For instance, distinguishing that ``heat'' signifies a temperature entity in RDF data remains problematic. 
Therefore, LLMs' proficiency in NLP tasks supports our approach to enhancing entity extraction and reducing manual and data-intensive dependencies \cite{Brown2020}, as further discussed in Section~\ref{sec:approach_overview}.

\paragraph{\textbf{Question-Answering over LD and RDF Triplestores.}} Traditional approaches to developing models for question-answering systems require specific training for each dataset or adjustments due to structural changes in the data. This process necessitates developing and maintaining models trained at comprehending and interpreting data, which is both resource-intensive and time-consuming due to the complexity and variability inherent in the data \cite{Dimitrakis2020}. As a result, conventional methods are impractical for creating general-purpose solutions that require adaptability, such as toolkits that can interface with multiple datasets or manage changes within the same dataset. In response to these challenges, the adoption of LLMs provides a promising alternative. These models enhance the adaptability and efficiency of question-answering systems by reducing the need for extensive retraining and facilitating more flexible integration with diverse and dynamically evolving datasets, such as those in observational LD \cite{Petroni2019,Brown2020,Hu2023,Tan2023}.

\paragraph{\textbf{LLMs Limitations.}} While LLMs exhibit significant capabilities, they have notable limitations that render them unsuitable for directly replacing chatbots employed within conversational UIs. These UIs, such as the one utilised in our experiment, are designed to automate the process of populating user selections on the UI, thereby constructing SPARQL queries to extract information from LD based on user input. One major issue is that LLM responses are not guaranteed to be consistently structured as expected, which can disrupt or completely break the UI interactivity, which poses a significant risk \cite{Zhang2022b,Hu2023,Ji2023}. Additionally, LLMs are prone to hallucinations, generating responses that may be plausible but are factually incorrect or irrelevant, further compromising the reliability of the interaction \cite{Ji2023}. To mitigate these issues, fine-tuning LLMs is often necessary to improve their ability to provide accurate and contextually appropriate responses to better align their output with the expected format and content, thereby restricting their responses to be more consistent and relevant to the task at hand \cite{Brown2020,Zhang2022b}. Fine-tuning can enhance the performance of LLMs in specialised applications, but these measures do not entirely eliminate the inherent risks associated with LLMs in contexts of system interaction \cite{Brown2020,Ji2023}. Section~\ref{sec:generate_sparql} explores another limitation of LLMs in generating SPARQL queries.

\section{Proposed Approach}
\label{sec:approach_overview}

\begin{figure}[h]
    \centering
    \includegraphics[width=0.8\textwidth]{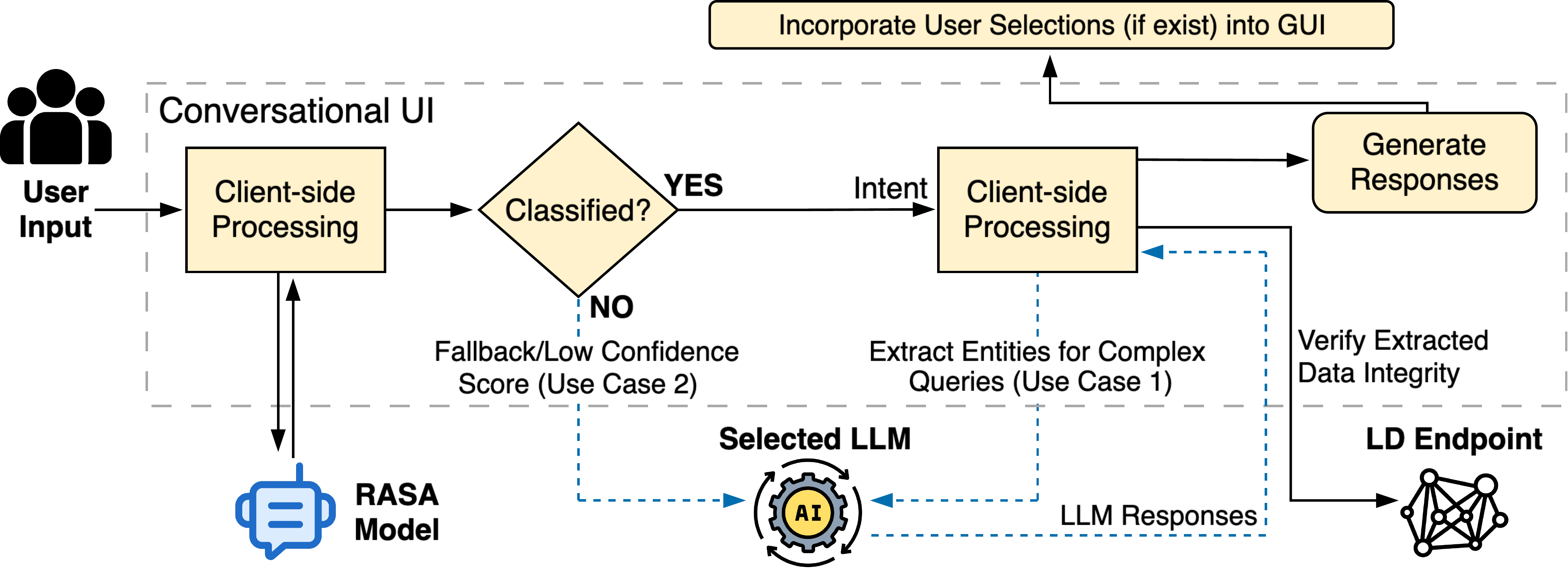}

    \caption{A conceptual representation of the conversational UI workflow within the current system. The blue dashed arrows illustrate the proposed enhancements to improve query formulation and information retrieval processes.}
    \label{fig:tool_overview}
\end{figure}

ForestQB~\cite{Mussa2024} is a novel SPARQL query builder that combines a chatbot interface with a form-based GUI to construct and execute SPARQL queries, enhancing the extraction of relevant information from observational LD; the source code is available at (\url{https://github.com/i3omar/ForestQB}). The integration of a chatbot has sped up the query construction, allowing users to seamlessly translate inquiries into the GUI (see Figure~\ref{fig:tool_overview}). Testing of ForestQB shows increased user satisfaction and a significantly reduced learning curve. However, the chatbot has limitations, notably in its inability to perform all tasks achievable by the form-based GUI, such as extracting and filtering an arbitrary set of entities from user queries. These limitations arise from a design choice to keep the chatbot ontology independent, ensuring compatibility with various datasets but restricting the complexity of questions and accuracy.


\begin{figure}[h]
    \centering
    \includegraphics[width=0.8\textwidth]{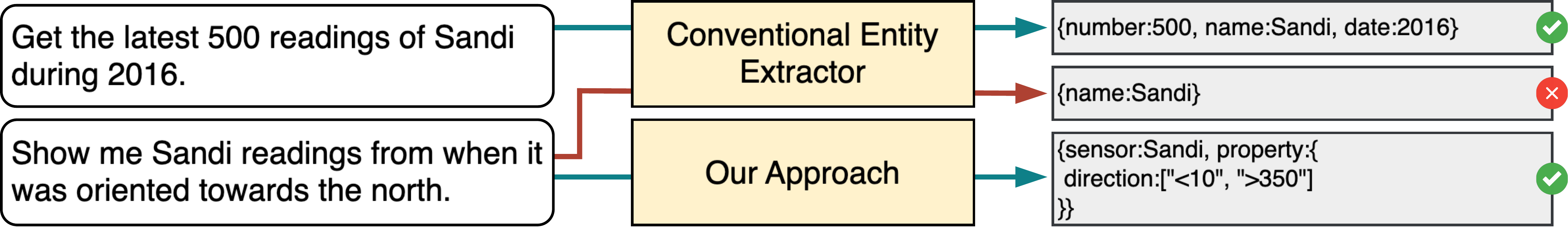}

     \caption{Illustration of Use Case 1: A scenario demonstrating the limitations of conventional entity extractors.}
    \label{fig:entity_extraction_example}
\end{figure}

To overcome these limitations, it is proposed that an LLM be integrated as an advanced entity extractor and reasoning tool, as shown in Figure~\ref{fig:tool_overview}. This improvement would enable the chatbot to directly utilise RDF data definitions and better respond to complex queries, particularly enhancing its application in wildlife ecology. Therefore, we propose two use cases to address these shortcomings and enhance the expressivity of the tool using LLMs as follows:

    \paragraph{\textbf{Use Case 1: Applying Filters Freely on the Sensor Properties using the Conversational UI}.} This refers to enabling users to apply all available filters within the graphical user interface freely using the chatbot, without requiring retraining to perform these tasks specifically (see Figure~\ref{fig:entity_extraction_example}).


    \paragraph{\textbf{Use Case 2: Querying RDF Data Schemas by using Conversational UI}.} This use case involves enabling users to query the sensors and data available within the RDF definitions of the dataset using the conversational UI. It should be general and adaptable to any observational LD endpoint we connect to, without necessitating training of the chatbot model to answer these specific questions.

\section{LLM Integration}

\subsection{Generating SPARQL query using LLMs}
\label{sec:generate_sparql}
We initially examined the use of LLMs to generate accurate SPARQL queries, encountering significant challenges \cite{Faria2023}. Our experiment with GPT-4 involved generating queries from a range of simple inputs specified by the toolkit, along with the required RDF definition file to guide the model. However, the results were largely unsuccessful; all generated queries failed to yield results except for the first, which was incorrect. The queries were structurally sound and adhered to SPARQL standards. However, the model reverted to its intrinsic understanding of the SOSA~\cite{Janowicz2019} ontology inherent to the dataset, thereby neglecting the specific context and data provided. Despite demonstrating precise encoding capabilities, effective entity extraction, and appropriate filter application, the findings suggest that LLMs are not yet reliable for directly converting user queries into SPARQL, though they may enhance data extraction methods. More details available at (\url{https://github.com/i3omar/LLM-Integration-Data}).

\subsection{RDF Embeddings}
Incorporating RDF data into the prompts of LLMs like GPT presents significant challenges primarily due to the voluminous nature of RDF datasets. The principal issue arises from the economic implications of embedding large RDF datasets into LLM prompts, especially with fee-based services such as OpenAI's GPT, where costs are proportional to the number of tokens processed. Additionally, although the maximum allowable prompt size has recently expanded, it remains insufficient for embedding substantial RDF data. For instance, models like Llama-3 accommodate a context length of up to 8192 tokens, which is considerably smaller than even a modest-sized RDF graph, particularly those containing observational data. To address these limitations, we have adopted the following approach:

    \paragraph{\textbf{1. Selective Data Inclusion:}} The definitions of sensors, their properties, and the corresponding properties each sensor monitors are the essential data for the LLM to identify entities and understand their relationships within the graph. Consequently, we omit less critical observational data from the LLM context. This selective inclusion significantly reduces the volume of RDF data while still transmitting valuable information. However, this does not imply that the entire data definition can be included, as it may still be excessively large.
    
\begin{figure}[h]
    \centering
    \includegraphics[width=0.7\textwidth]{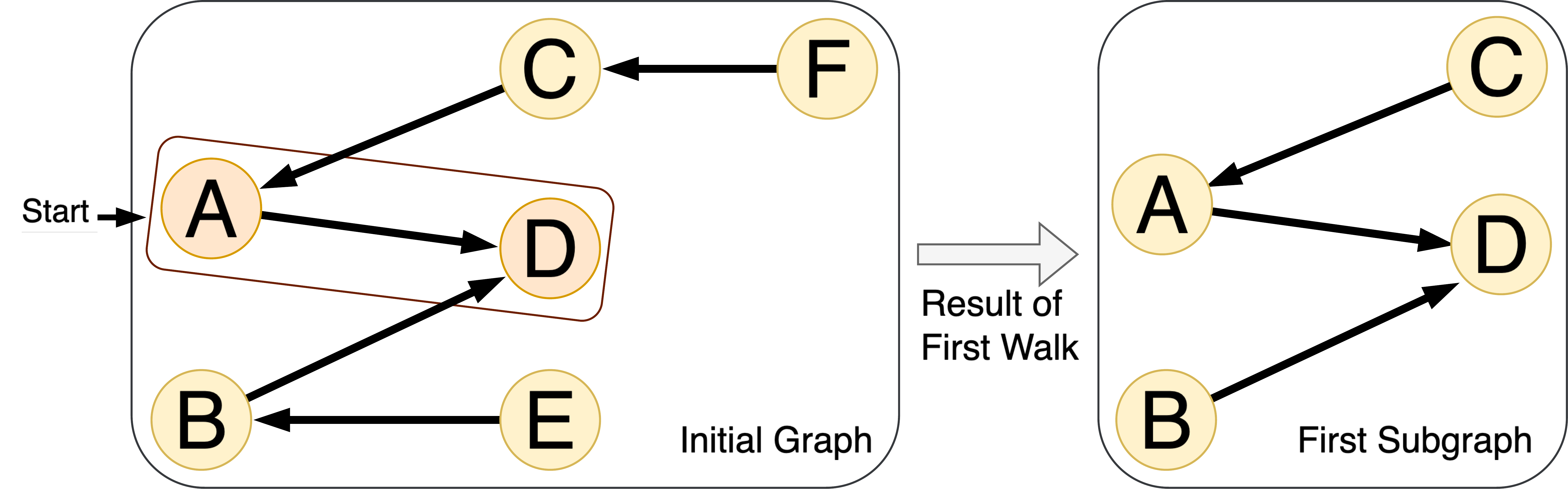}

    \caption{An example of the first iteration of RDF Walk showing the first generated subgraph.}
    \label{fig:rdf_walks}
\end{figure}

    \paragraph{\textbf{2. Subgraph Generation through RDF Walks:}} Unlike document embeddings, which are tailored for unstructured text, RDF embeddings are specifically designed to manage structured graph data. A viable approach involves decomposing extensive RDF graphs into smaller, more manageable subgraphs. This decomposition is achieved by using RDF walks, inspired by the RDF2Vec technique \cite{Ristoski2016}, to create subgraphs that encapsulate each triple's directly connected nodes. Figure \ref{fig:rdf_walks} illustrates the first subgraph resulting from the initial walk. However, RDF2Vec was found to be impractical for our specific applications due to the high computational demands associated with dynamically generating subgraphs from user queries. For example, take the query, `What is the latest reading of Sensor A?' Within this query, the identifiable entities are `Sensor A' and `latest reading.' Subsequently, it becomes imperative to construct a simplified RDF graph or subgraph that accurately encapsulates the semantic structure of the user query. Instead, we decompose the RDF graph into individual triples stored in memory as an array. These triples are then iteratively processed to identify and amalgamate directly connected triples into smaller subgraph arrays.

\begin{figure}[h]
    \centering
    \includegraphics[width=0.8\textwidth]{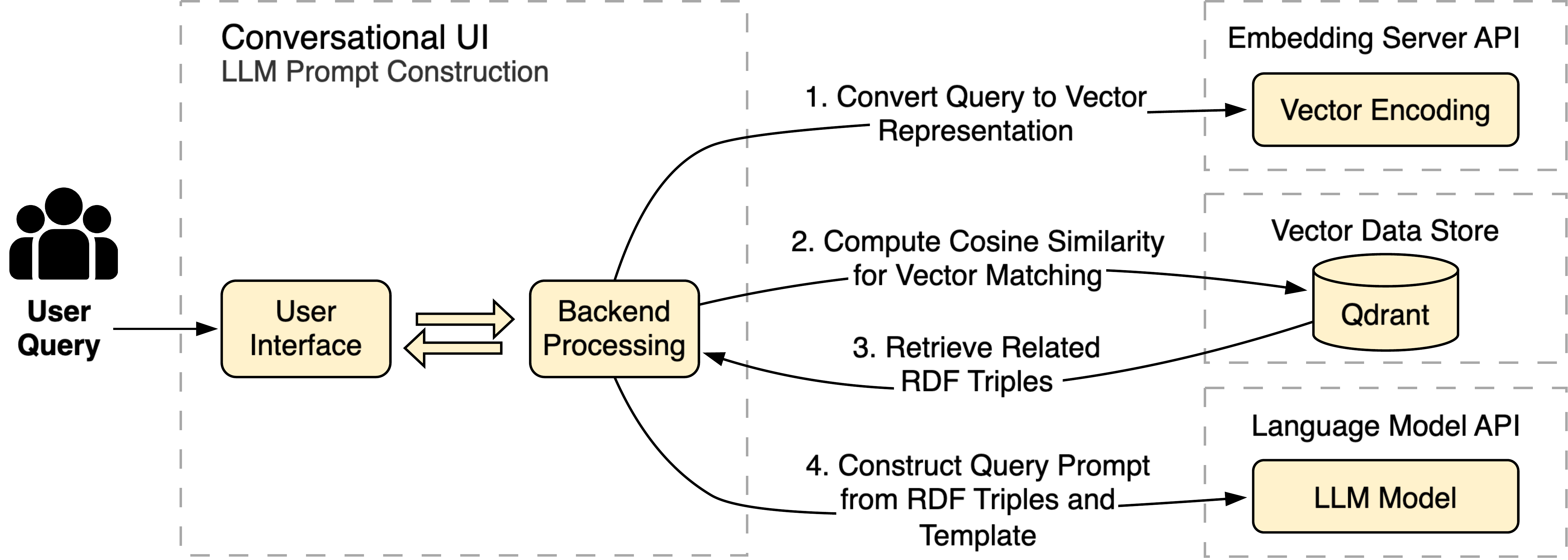}

    \caption{A summary of the workflow for generating LLM prompts: It begins with the conversion of user queries into vector representations, followed by similarity calculations to identify relevant RDF triples. These triples form a subgraph that, combined with a prompt template, aids in producing context-aware prompts.}
    
    \label{fig:embedding_approach}
\end{figure}

    \paragraph{\textbf{3. Vector Embedding of RDF Triples:}} As each RDF triple selected in this step is enriched with textual annotations, including labels and descriptive comments specific to the entity, we utilised the `paraphrase-TinyBERT-L6-v2', a pre-trained model capable of capturing semantic similarities within textual data \cite{Reimers2019}, to represent each triple as a vector. These vectors are then stored in Qdrant, an open-source vector database. As summarised in Figure~\ref{fig:embedding_approach}, when a user query is received, it is similarly encoded, and a cosine similarity search is performed in Qdrant to identify the most closely related triple. Some entities in the questions are associated with disjoint subgraphs. By merging the highest-scoring entities, we significantly increase the likelihood of including all essential data in the injected RDF. Therefore, the two highest-scoring triples are identified, and a union of their corresponding subgraphs is retrieved. These subgraphs are then incorporated into the LLM prompt to ensure that all relevant data is included.
    

    The selection of this particular model was motivated by its ability to identify all related triples for the queries associated with the experiment for use cases 1 and 2. We conducted evaluations on multiple state-of-the-art pre-trained sentence embedding models available on Hugging Face, utilising the SentenceTransformers framework. These evaluations involved conducting semantic searches using Qdrant with the vector representations derived from these models. Typically, we observed an enhancement in the average accuracy across all models by 13.15\%, achieved through the integration of subgraphs from the two highest-ranked indices in the similarity searches. This approach culminated in attaining 100\% accuracy with our preferred model, as detailed in Table~\ref{tab:embeddings_result_table}.

\begin{table}[]
  \centering
  \caption{Comparison of similarity search results using different pretrained embedding models, detailing the accuracy achieved when using only the highest scoring index and when incorporating the top two highest scoring indices.}
    \begin{NiceTabular}{lcc}
    \toprule
    Model Name& \multicolumn{1}{c}{Top Index} & \multicolumn{1}{c}{Top 2 Indices} \\
    \hline
    paraphrase-TinyBERT-L6-v2~\cite{Reimers2019} & \textbf{86\%}  & \textbf{100\%} \\
    multi-qa-mpnet-base-cos-v1 & 78\%  & 90\% \\
    all-MiniLM-L6-v2 & 69\%  & 86\% \\
    multi-qa-mpnet-base-dot-v1 & 79\%  & 85\% \\
    hkunlp/instructor-large & 75\%  & 85\% \\
    sentence-t5-base~\cite{Ni2021} & 73\%  & 83\% \\
    all-mpnet-base-v2 & 65\%  & 81\% \\
    multi-qa-MiniLM-L6-cos-v1 & 73\%  & 80\% \\
    \bottomrule
    \end{NiceTabular}%
  \label{tab:embeddings_result_table}%
\end{table}%

\subsection{Prompts templates}
An essential step in enhancing the accuracy of outputs generated by LLMs is the preparation of an appropriate prompt. Recent studies have demonstrated the high sensitivity of LLMs to the specificity of prompts \cite{Zhang2022,Liu2022,Agrawal2013}. Consequently, to facilitate the generation of valid JSON from user queries, a JSON template has been embedded within the prompt to provide the LLM with a clear indication of the expected output. In addition, two distinct types of prompts were developed to evaluate the most effective method for extracting entities using zero-shot examples. The first prompt was succinct, offering minimal instruction, while the second was more elaborate, clarifying relevant filters to assess whether this would lead to enhanced output accuracy. We conducted experiments using `gpt-3.5-turbo-0125' on a set of 20 questions, equally divided between 10 straightforward and 10 complex inquiries, to ascertain whether the outputs differed significantly between the two prompts. As shown in Table~\ref{tab:prompt_testing}, the results indicated a 42.86\% improvement in the outputs generated by the more detailed second prompt compared to the first. This finding suggests that more comprehensive instructions significantly aid the model in producing more accurate outputs. Therefore, the more detailed prompt will be utilised in our subsequent experiments.


\begin{table}[]
  \centering
  \caption{Comparison of accuracy and improvement percentages between two LLM prompts across three evaluation tasks: Extracting Entities, Extracting Properties, and Applying Filters, with overall success in all tasks.}
    \begin{tabular}{lccc}
    Task  & Prompt-1 & Prompt-2 & Improvement \\
    \hline
    Extracting Entities & 90\% & 95\% & 5.56\% \\
    Extracting Properties & 65\% & 85\% & 30.77\% \\
    Applying Filters & 35\% & 50\% & 42.86\% \\
    \hline
    Full Task Success & 35\% & 50\% & \textbf{42.86}\% \\
    \hline
    \end{tabular}%
  \label{tab:prompt_testing}%
\end{table}%

\section{Experiment Evaluation}
In this section, the effectiveness of incorporating LLMs into the existing system to improve information retrieval is evaluated, alongside an analysis of the experimental results. The results and evaluation metrics are available at (\url{https://github.com/i3omar/LLM-Integration-Data}).

\subsection{Dataset}
We conducted the experiment using a private SPARQL endpoint hosting observational LD, structured in accordance with the SOSA~\cite{Janowicz2019} ontology. The dataset is publicly available and can be accessed through the GitHub repository at (\url{https://github.com/i3omar/ForestRDF}).

\subsection{Selected Large Language Models}
In order to rigorously evaluate our approach, we have selected a diverse range of state-of-the-art language models. Our selection primarily includes six versions of OpenAI's latest GPT-3.5 and GPT-4 models (\url{https://platform.openai.com/docs/models}). Additionally, we integrate two open-source models into our study: openchat-3.5-0106~\cite{Wang2023} and Meta-Llama-3-8B-Instruct~\cite{Llama3}. The former has been finetuned from the `Mistral-7B-v0.1' base model, while the latter is the most recent iteration within the Meta Llama-3 series. Both open-source models were chosen due to their lightweight architecture, making them practical for real-world applications. Furthermore, all selected models have demonstrated superior performance in various NLP tasks relevant to our study. The primary objective of this selection is not to compare these models directly; rather, it is to validate our approach's effectiveness and achieve high accuracy in its application.

\subsection{Experiment Setup}
We have formulated a set of 40 questions designed to assess the application of LLMs across use cases 1 and 2, where the LLM was prompted with relevant RDF data. These questions are strategically divided into four groups to gauge the effectiveness of the LLM model in various scenarios:

\begin{enumerate}
    \item \textbf{Simple Direct Queries}: Involves a single entity clearly identified in the RDF data.
    \item \textbf{Complex Direct Queries}: Encompasses questions with multiple entities, all explicitly mentioned in the RDF data.
    \item \textbf{Simple Indirect Queries}: Queries about a single entity not directly mentioned in the data, using synonyms and different phrasing to evaluate the model's inference abilities.
    \item \textbf{Complex Indirect Queries}: More involved queries that include multiple entities not explicitly stated, utilising synonyms and varied wording to assess the model's comprehension and reasoning capabilities.
\end{enumerate}

We conducted six trials per query, equally split between zero-shot and few-shot tests, to evaluate output consistency and the effect of examples on model performance. Each model underwent 120 queries in both zero-shot and few-shot settings, totaling 240 queries. However, for Use Case 2, only zero-shot testing was performed, totalling 120 queries.

\subsection{Evaluation Metrics}
To evaluate our approach, we conducted manual evaluations for both use cases and automatic validation of the JSON output for Use Case 1, applying targeted metrics to ensure practical and technical validity, as follows:


\subsubsection{A. Manual Evaluation (Accuracy)}
The effectiveness of the LLM in enhancing entity extraction and RDF reasoning capabilities was manually evaluated through meticulous testing in practical scenarios. The output is manually classified in a binary manner as correct or incorrect. For Use Case 1, the output is deemed correct if the generated JSON output includes the correct answer and does not exhibit issues such as incorrect key naming or invalid JSON format. Meanwhile, for Use Case 2, the output comprises plain text that has been manually evaluated against reference data to verify the accuracy of the generated output.


\subsubsection{B. Automatic Structural Evaluation}

In practical settings for Use Case 1, two models can generate outputs that are both correct and compatible with the system, although one model may introduce additional keys (which remain operational). This evaluation aims to determine the extent to which each model complies with specified instructions to yield a more reliable JSON output. These adapted metrics, based on foundational work in the field as detailed by~\cite{Manning2008,Reimers2019}, evaluate performance by comparing JSON outputs from LLMs to a reference answer, enabling a deeper examination of quality beyond mere accuracy. The metrics used are as follows:


\paragraph{\textbf{B.1. Structural Similarity Score (StrSS):}} This metric measures the similarity between the structure of the reference JSON and the generated JSON output. Both JSON objects are flattened, with each key representing a path through the original structure. URIs in the generated JSON are simplified to their abbreviated forms used in the reference, e.g., converting `http://.../sosa/resultTime' to `sosa:resultTime' to prevent scoring penalties for acceptable URI variations. Extra keys in the output incur a lesser penalty, recognising that models may generate correct but additional keys, slightly lowering the match score for a more nuanced comparison. The metric equation for evaluating outputs is as follows:

\begin{equation}
\text{StrSS} = \frac{|K_{\text{ref}} \cap K_{\text{gen}}|}{|K_{\text{ref}}| + \beta \cdot (|K_{\text{gen}} - K_{\text{ref}}|)}
\end{equation}

\noindent where \( K_{\text{ref}} \) and \( K_{\text{gen}} \) denote the sets of keys in the reference and generated JSON objects, respectively. The term \( |K_{\text{ref}} \cap K_{\text{gen}}| \) quantifies the intersection, representing keys common to both sets. Conversely, \( |K_{\text{gen}} - K_{\text{ref}}| \) measures the keys unique to \( K_{\text{gen}} \). The weighting factor \( \beta = 0.1 \) moderates the effect of these unique keys in the scoring mechanism, reducing their impact.

\paragraph{\textbf{B.2. Semantic Similarity Score (SemSS):}} This metric evaluates the semantic similarity between values in reference and generated JSONs by converting them into vector embeddings using the `paraphrase-TinyBERT-L6-v2' model. Cosine similarity between these embeddings measures their semantic closeness. The average cosine similarity across corresponding key-value pairs in the JSON structures quantifies semantic alignment, defining the SemSS score. The SemSS score is defined as:


\begin{equation}
\text{SemSS} = 
\begin{cases} 
\frac{1}{N} \sum_{i=1}^{N} \text{cos\_sim}\left(\text{emb}(v_{\text{ref}, i}), \text{emb}(v_{\text{gen}, i})\right) & \text{if } N > 0 \\
0 & \text{otherwise}
\end{cases}
\end{equation}

\noindent where $N$ is the count of keys shared between \( K_{\text{ref}} \) and \( K_{\text{gen}} \). For the $i$-th matching key, $v_{\text{ref}, i}$ and $v_{\text{gen}, i}$ denote the associated values in the reference and generated JSONs, respectively. The function $\text{emb}(v)$ computes the embedding of value $v$, and $\text{cos\_sim}(a, b)$ calculates the cosine similarity between vectors $a$ and $b$.


\paragraph{\textbf{B.3. Overall JSON Similarity Score (OJSS):}} This score is a weighted average that combines the two distinct similarity scores: the SemSS Score and the StrSS Score. Both scores are assigned equal weight, each contributing 50\% to the final calculation, denoted by a coefficient of 0.5 for each. This equal weighting underscores the balanced importance of both the structural integrity of the keys and the semantic accuracy of the values in determining the overall score. The following is the OJSS complete equation:

\begin{equation}
\text{OJSS} = \alpha \times \text{StrSS} + (1 - \alpha) \times \text{SemSS}
\end{equation}
\noindent where $\alpha$ = 0.5, $\alpha$ is a weighting factor between 0 and 1 that balances the importance of structural and semantic similarity.

\subsection{Use Case 1: Results}

\begin{figure}[h]
    \centering
    \includegraphics[width=\textwidth]{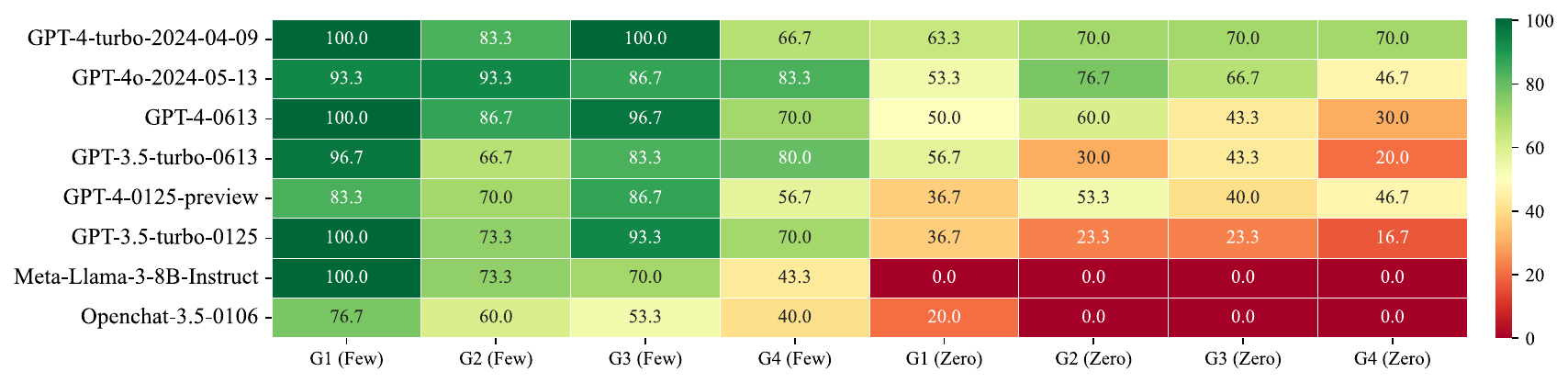}

    \caption{Accuracy of selected LLMs for Use Case 1, categorised by question group (G) and test type (few-shot or zero-shot). Green indicates higher accuracy, while red shows lower accuracy.}
    \label{fig:heatmap_uc1}
\end{figure}

Table \ref{tab:usecase1_result} presents the performance of the selected LLMs in generating accurate JSON outputs, reflecting the queries employed in our research. This practical experiment relied on the system's capability to retrieve and process the generated responses accurately and to construct the queries correctly to be deemed successful. Figure~\ref{fig:heatmap_uc1} illustrates responses across question groups.


\begin{table}[b!]
\centering

\caption{Performance metrics comparison of LLMs in Use Case 1 for few-shot and zero-shot scenarios, showing percentages for accuracy, structural similarity (StrSS\%), semantic similarity (SemSS\%), and overall JSON similarity (OJSS\%). Sorted by overall accuracy (OA\%) in descending order.}
\label{tab:usecase1_result}
\resizebox{\textwidth}{!}{
\begin{tabular}{lccccc|ccccc|c}
    \toprule
    & \multicolumn{5}{c}{Few-shot} & \multicolumn{5}{c}{Zero-shot} \\
    \cmidrule(lr){2-6} \cmidrule(lr){7-11}
      Model Name & \multicolumn{2}{c}{Accuracy}  & \multicolumn{1}{c}{StrSS} & \multicolumn{1}{c}{SemSS} & \multicolumn{1}{c}{OJSS} & \multicolumn{2}{c}{Accuracy} & \multicolumn{1}{c}{StrSS} & \multicolumn{1}{c}{SemSS} & \multicolumn{1}{c}{OJSS} & \multicolumn{1}{c}{OA}\\
    \midrule
    GPT-4-turbo-2024-04-09 & 87.5 & $\pm 0.03$  & \textbf{87.4} & \textbf{84.2} & \textbf{85.8} & \textbf{68.3} & $\pm 0.04$  & \textbf{71.9} & \textbf{63.2} & \textbf{67.6} & \textbf{77.92} \\
    GPT-4o-2024-05-13 & \textbf{89.2} & $\pm 0.06$  & 83.0 & 79.6 & 81.3 & 60.8 & $\pm 0.08$  & 69.0 & 60.0 & 64.5 & 75.00 \\
    GPT-4-0613 & 88.3  & $\pm 0.03$  &  60.8 & 63.2 & 62.0 & 45.8 & $\pm 0.06$  & 35.6 & 34.5 & 35.0 & 67.08 \\
    GPT-3.5-turbo-0613 & 81.7 & $\pm 0.11$  & 53.1 & 56.1 & 54.6 & 37.5 & $\pm 0.09$  & 4.5 & 4.4 & 4.5 & 59.58 \\
    GPT-4-0125-preview & 74.2 & $\pm 0.07$  & 58.4 & 59.9 & 59.2 & 44.2 & $\pm 0.06$  & 30.3 & 25.8 & 28.1 & 59.17 \\
    GPT-3.5-turbo-0125 & 84.2 & $\pm 0.07$  & 56.6 & 56.5 & 56.5 & 25.0 & $\pm 0.09$  & 28.2 & 25.6 & 26.9 & 54.58 \\
    Meta-Llama-3-8B-Instruct  & 71.7 & $\pm 0.05$  & 58.8 & 57.7 & 58.3 & 0.0 & $\pm 0.00$  & 0.0 & 0.0 & 0.0 & 35.83 \\
    Openchat-3.5-0106 & 57.5 & $\pm 0.05$  & 55.8 & 53.9 & 54.9 & 5.0 & $\pm 0.00$  & 24.3 & 17.2 & 20.8 & 31.25 \\
    \bottomrule
\end{tabular}}
\end{table}

\paragraph{Zero-shot} During the zero-shot testing phase, the performance levels were notably low as most LLMs did not adhere precisely to the instructions. For instance, Llama-3, despite explicit instructions to generate solely JSON outputs, produced additional explanatory text alongside the JSON, thereby rendering the outputs impractical for further processing, culminating in a complete failure rate. In contrast, GPT-4-turbo achieved an accuracy of 68.3\%, with a StrSS score of 71.9\%, indicating a well-structured and highly practical output.

\paragraph{Few-shot} In the few-shot tests, there was a substantial improvement in results. LLMs utilised the provided examples as historical references, enhancing their ability to generate structured outputs, as evidenced by the StrSS scores. GPT-4o achieved the highest accuracy rate at 89.2\%, followed by GPT-4-0613 and GPT-4-turbo, with scores of 88.3\% and 87.5\% respectively. GPT-4-turbo also scored 87.4\% in StrSS, 84.2\% in SemSS, and 85.8\% in OJSS, demonstrating fewer errors in practical applications within web information systems, despite not being the most accurate model. Furthermore, Llama-3 showed a notable increase in accuracy to 71.7\%, a commendable achievement for a lightweight model, illustrating greater adherence to instructions when provided with few-shot examples.


\subsection{Use Case 2: Results}

\begin{figure}[h]
    \centering
    \includegraphics[width=0.8\textwidth]{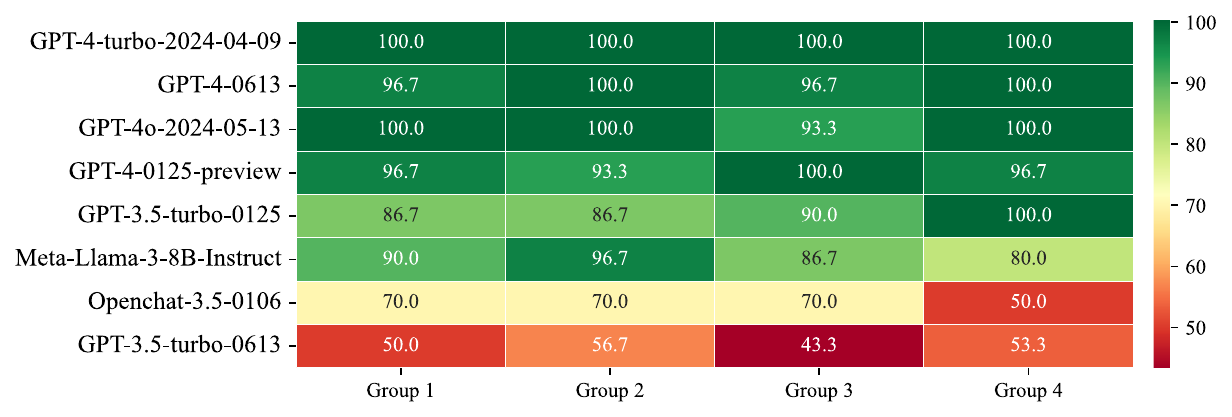}
        \caption{Accuracy of selected LLMs for Use Case 2 questions, categorised by group number. Higher accuracy is indicated in green, while lower accuracy is shown in red.}
        
    \label{fig:heatmap_uc2}
\end{figure}

The application of LLMs for reasoning over RDF data to respond to queries related to the data schema has demonstrated notable efficacy. By using the GPT-4-turbo model, we achieved a remarkable accuracy rate of 100\% (see Table~\ref{tab:usecase2_result}). As seen in Figure~\ref{fig:heatmap_uc2}, the performance of most models, including open-source ones, was generally robust; however, the `GPT-3.5-turbo-0613' model underperformed significantly due to its limitations in processing RDF data as textual content, reflecting its lack of training for such data types. 

\begin{table}[hb!]
  \centering
  \caption{Comparative performance metrics of LLMs in Use Case 2, showing percentages for accuracy and responses containing RDF snippets (Snippet).}
  \resizebox{0.5\textwidth}{!}{

    \begin{NiceTabular}{lccc}
              \toprule

    Model Name & \multicolumn{2}{c}{Accuracy(\%)} & Snippet(\%) \\
              \hline

    GPT-4-turbo-2024-04-09 & \textbf{100.0} & $\pm 0.00$ & 0.00 \\
    GPT-4-0613 & 98.33 & $\pm 0.03$  & 1.67 \\
    GPT-4o-2024-05-13 & 98.33 & $\pm 0.01$  & 20.83 \\
    GPT-4-0125-preview & 96.67 & $\pm 0.04$  & 4.17 \\
    GPT-3.5-turbo-0125 & 90.83 & $\pm 0.03$  & 0.00 \\
    Meta-Llama-3-8B-Instruct  & 88.33 & $\pm 0.03$  & 24.17 \\
    Openchat-3.5-0106 & 73.33 & $\pm 0.10$  & 3.33 \\
    GPT-3.5-turbo-0613 & 50.83 & $\pm 0.44$  & 0.00 \\
              \bottomrule
    \end{NiceTabular}%
    }
  \label{tab:usecase2_result}%
\end{table}%

The capacity of these models to interpret RDF relationships varies; for instance, the Openchat model exhibited some reasoning abilities but struggled with comparisons between RDF entities and occasionally generated responses not present in the RDF data. In contrast, the Llama-3 model demonstrated more substantial reasoning capabilities, which is notable for its smaller size. Overall, GPT-4 models have shown significant improvements in reasoning over RDF data, suggesting a higher level of training for such tasks.

Moreover, we observed that some models embed RDF snippets in their responses, reducing readability and diminishing accessibility, particularly when the data is meant to remain hidden from the user, thus adding to the confusion. As shown in Table~\ref{tab:usecase2_result}, the highest rates of RDF snippet inclusion were observed in the responses from Llama-3 and GPT-4o, amounting to 24.17\% and 20.83\% respectively.

\subsection{Result Analysis}
While a majority of the models demonstrated competent performance in zero-shot scenarios, particularly with Use Case 2, the challenges of executing Use Case~1 in similar settings were pronounced. In the analysis of error patterns within Use Case 1 across various LLMs, multiple shortcomings became evident. Notably, models frequently misinterpreted `Horizontal Dilution of Precision (HDOP)' for `Positional Dilution of Precision (PDOP)', particularly observable in tasks 3.10 and 4.4. These tasks relied on LLM RDF reasoning capabilities to identify the correct entity, complicated by the similarity in their descriptions. Similarly, during task 4.7, certain LLMs interpreted the term `not empty' as `not bound' despite explicit instructions to the contrary within the prompt, thus revealing inconsistencies in adhering to stipulated guidelines. Moreover, task 4.1 underscored a confusion between `activity' and `speed', underscoring limited abilities in interpreting RDF schemas in some models. Additionally, tasks 1.7 and 3.7 highlighted variability in estimating `north degree', where structurally correct responses with nearly accurate angle values were deemed acceptable, given that users could manually adjust inaccuracies.

Most of these errors were significantly mitigated by introducing a few-shot learning approach, which facilitated a deeper understanding of the tasks by the models. As illustrated in Figure~\ref{fig:heatmap_uc1}, the results markedly improved across the same set of queries (groups) when tested in few-shot scenarios. Thus, within practical settings aimed at augmenting entity extraction and RDF data reasoning, LLMs prove to be effective and can enhance the expressivity of conversational UIs to interpret user responses more accurately.

\section{Conclusion}
In this paper, we propose a technique for integrating LLMs into existing conversational UIs, enhancing their functionality without requiring retraining. Our approach leverages the intrinsic strengths of LLMs to complement traditional chatbot models and improve entity extraction over   LD and RDF triplestores. We initiated our discussion by outlining the system's limitations with the proposed use cases, followed by a critical analysis of why direct SPARQL generation or outright replacement of the existing chatbot model with LLMs was suboptimal. Our proposed solution includes embedding RDF schema into LLM prompts to ensure the question is contextually enriched with relevant RDF data. Furthermore, we enhanced the quality of LLM responses through optimised prompt templates, leading to more accurate and context-aware interactions. Our experimental evaluations corroborate that LLMs can effectively function as both an entity extraction component and a reasoning tool within our system, significantly augmenting user experience and improving the overall information retrieval process. This integration addresses key limitations of prior systems, such as scalability and contextual understanding, and sets a new benchmark for advancements in conversational UIs and LD retrieval. Our approach, enhancing semantic query handling and entity recognition, paves the way for further research into LLM applications across diverse datasets and domains.

\subsubsection{Acknowledgements.} We would like to acknowledge the scholarship and continuous support provided by Saudi Electronic University. This work was also partly supported by the HEFCW ODA Awards, Danau Girang Field Centre (DGFC), EPSRC Institutional Sponsorship, GCRF Facilitation funding, and ARCCA at Cardiff University.

\bibliographystyle{splncs04}
\bibliography{references}

\end{document}